\documentclass[twocolumn,
showpacs,
amsmath,
amssymb,
amsfonts,
final,
a4paper,
aps,
citeautoscript,
footinbib,
pra,
superscriptaddress]{revtex4}
\usepackage{times}
\usepackage{graphicx}
\usepackage[latin1]{inputenc}

\graphicspath{{./Figs/}}
\def\Ham{\mathcal{H}}

\newcommand{\ket}[1]{\left\vert{#1}\right\rangle}
\newcommand{\bra}[1]{\left\langle{#1}\right\vert}
\newcommand{\md}[1]{\vert{#1}\vert}
\newcommand{\Md}[1]{\left\vert{#1}\right\vert}

\newcommand{\moy}[1]{\langle{#1}\rangle}
\newcommand{\inter}[2]{\langle{#1}\vert{#2}\rangle}
\newcommand{\elem}[3]{\langle{#1}\vert{#2}\vert{#3}\rangle}

\begin{document}

\author{Guillaume Roux}
\email{guillaume.roux@u-psud.fr}
\affiliation{LPTMS, CNRS and Universite Paris-Sud, UMR8626, Bat. 100,
  91405 Orsay, France.}

\date{\today} 

\title{Finite size effects in global quantum quenches: examples from
  free bosons in an harmonic trap and the one-dimensional Bose-Hubbard
  model}


\begin{abstract}
  We investigate finite size effects in quantum quenches on the basis
  of simple energetic arguments. Distinguishing between the low-energy
  part of the excitation spectrum, below a microscopic energy-scale,
  and the high-energy regime enables one to define a crossover number
  of particles that is shown to diverge in the small quench
  limit. Another crossover number is proposed based on the fidelity
  between the initial and final ground-states. Both criteria can be
  computed using ground-state techniques that work for larger system
  sizes than full spectrum diagonalization. As examples, two models
  are studied: one with free bosons in an harmonic trap which
  frequency is quenched, and the one-dimensional Bose-Hubbard model,
  that is known to be non-integrable and for which recent studies have
  uncovered remarkable non-equilibrium behaviors. The diagonal weights
  of the time-averaged density-matrix are computed and observables
  obtained from this diagonal ensemble are compared with the ones from
  statistical ensembles. It is argued that the ``thermalized'' regime
  of the Bose-Hubbard model, previously observed in the small quench
  regime, experiences strong finite size effects that render difficult
  a thorough comparison with statistical ensembles. In addition, we
  show that the non-thermalized regime, emerging on \emph{finite size
  systems} and for \emph{large} interaction quenches, is not related
  to the existence of an equilibrium quantum critical point but to the
  high energy structure of the energy spectrum in the atomic
  limit. Its features are reminiscent of the quench from the
  non-interacting limit to the atomic limit.
\end{abstract}

\maketitle 

The study of the non-equilibrium evolution of closed quantum many-body
systems has been triggered by the recent progresses in cold atoms
experiments in which atoms are hardly coupled to the
environment~\cite{Greiner2002, Experiments}. Furthermore, microscopic
parameters of the Hamiltonian governing the dynamics can be controlled
at will and changed on microscopic timescales. In this context, the
question of the unitary evolution of an isolated quantum system after
a sudden change of one parameter, the so-called quantum quench, has
attracted a lot of interest in both the experimental and theoretical
communities~\cite{Moeckel2009}. Many different questions are raised by
such a set-up, among which are the relaxation of
observables~\cite{Igloi2000, Sengupta2004, Rigol2006, Cazalilla2006,
  Perfetto2006, Calabrese2006, Calabrese2007, Kollath2007,
  Manmana2007, Gangardt2008, Hackl2008, Babadi2009, Hackl2009,
  Barthel2009}, the question of thermalization~\cite{Peres1984,
  EarlyETH, Deutsch1991, Srednicki1994, Srednicki1994a, Srednicki1999,
  Skrovseth2006, Rigol2007, Kollath2007, Brody2007, Rigol2008,
  Reimann2008, Moeckel2008, Rossini2009, Eckstein2009, Rigol2009,
  Rigol2009a, Sotiriadis2009}, the existence of a subsystem steady-state~\cite{Barthel2008, Cramer2008,
  Cramer2008a, Flesch2008}, and the propagation of the
entanglement~\cite{Calabrese2005, Chiara2006, Lauchli2008,
  Fagotti2008, Manmana2009}. Beyond these academic concerns, practical
applications of quenches have been proposed through the engineering of
metastable states~\cite{Heidrich-Meisner2008, Heidrich-Meisner2009}
and of an out-of-equilibrium supersolid state in a cold atoms
set-up~\cite{Keilmann2009}. This paper is dedicated to the
thermalization issue, but restricted to specific examples and without
claims on general results about the thermalization mechanism. In this
context, a quench can be understood as a way to create an initial
state that evolves through the dynamics of a given Hamiltonian. A
common wisdom in classical mechanics is that the long-time evolution
will forget about the initial state and will explore all the
accessible phase-space, provided the dynamics are chaotic. Then,
ergodicity allows for the use of statistical ensembles in place of
time-averaging. For a closed quantum system, as the evolution is
unitary and the spectrum discrete, long-time recurrences occur and the
contribution of the eigenstates involved in the dynamics is fixed
by the initial state. For large enough systems, a quantum ergodic
theorem was proposed~\cite{Neumann1929}, supporting the emergence of
the microcanonical ensemble which is the usual statistical ensemble
for an isolated system. This approach aims at showing that
time-averaged density-matrix $\bar{\rho}$ (see below for the
definition) is macroscopically equivalent to the microcanonical
ensemble. In a quantum quench, the initial state is not a ``typical''
state of a given energy but usually is the ground-state of the same
Hamiltonian with different parameters. Consequently, the quench amplitude, or
how much do we change the Hamiltonian, is here another relevant
quantity. Another way to regard a quench can be as a perturbation of
the initial state and one may wonder whether the long-time response is
sensitive to the initial state. Furthermore, numerical tools and
experiments on closed systems cannot easily reach a large number of
particles so finite size effects can be important in the
interpretation of the observed phenomena. This paper suggests possible
approaches to the question of these finite size effects after a
quantum quench, and a possible interpretation of the observations made
on a particular model: the one-dimensional Bose-Hubbard model. The
other model, consisting of free bosons in an harmonic trap, offers
another example of finite size effects and remarkable behaviors. Some
of the features of the two models are surprisingly connected.

The central object governing the long-time physics after a quantum
quench is the time-averaged density-matrix $\bar{\rho}$ that predicts
the time-averaged expectation values of any observable. This
density-matrix has also connections to the heat or work done on a
system~\cite{Silva2008, Dorosz2008, Barankov2008, Polkovnikov2008c}.
The weights of this ``diagonal ensemble'' are difficult to compute for
large systems as one needs to fully diagonalize the Hamiltonian so one
unfortunately has to work with small systems (Hilbert spaces). Other
methods have been used to tackle the physics of quenches. For
instance, ``Ab-initio'' numerics have been used on both integrable and
non-integrable models~\cite{EarlyETH, Kollath2007, Manmana2007, Rigol2008,
  Eckstein2008, Kollar2008, Barthel2008, Cramer2008a, Flesch2008,
  Roux2009, Biroli2009}. Numerical methods like time-dependent
density-matrix renormalization group (tDMRG)~\cite{Vidal2004,
  White2004, Daley2004} can be used to compute the time-evolution of
the wave-function but the interpretation is restricted to observables
and to a finite window of time, and cannot give access the these
weights. Exact results on integrable models~\cite{Igloi2000,
  Rigol2007, Barthel2008, Cramer2008a, Flesch2008, Barmettler2009,
  Faribault2009, Li2009, Biroli2009} have the advantage to treat in a
non-perturbative way large systems, but on the other hand, it is not
surprising that they do not always thermalize due to the extensive
number of conserved quantities. Luttinger liquid theory, which
describes the low-energy physics of one-dimensional models in terms of
free bosonic fields (thus an integrable theory), has been used to
compute the time-evolution of the observables~\cite{Calabrese2006,
  Cazalilla2006, Perfetto2006, Iucci2009, Uhrig2009}. Quantum chaos
methods have also helped studying the time-evolution of the
Bose-Hubbard model~\cite{Bodyfelt2007, Cassidy2009, Hiller2009}. Some
studies focused on the relation between fidelity and on the energy
distribution~\cite{Roux2009,Zanardi}. All these methods suffer from
approximations and/or finite size effects and it is sometimes hard to
determine what is an artifact or not.

Some of the results from numerical simulations seems to be
contradictory~\cite{Kollath2007, Manmana2007, Rigol2008, Roux2009,
  Rigol2009, Rigol2009a} but were carried out on different models with
different range of parameters, and not necessarily starting from the
ground-state~\cite{Rigol2008} of a simply related
Hamiltonian. Performing a quantum quench amounts to projecting an
initial state onto the energy spectrum of the final Hamiltonian,
corresponding to a certain distribution of energy $\bar{\rho}(E)$. In
the thermodynamical limit, a global quench is expected to drive the
mean energy to the bulk of the energy spectrum since the perturbing
operator is extensive. In this high energy domain, semi-classical
physics and random-matrix theory arguments are expected to work and
make expectation values hardly depend on the energy (within a window
given by the energy fluctuations)~\cite{Peres1984, Deutsch1991,
  Srednicki1994}: thermalization can occur in the sense that the
energy distribution obtained from the quench gives the same averages
for the observables as the microcanonical ensemble. This so-called
``eigenstates thermalization hypothesis'' (ETH) has been tested
numerically~\cite{EarlyETH, Rigol2008, Rigol2009, Rigol2009a} for
given models (typically fermionic and hard-core bosonic models) and
some given set of parameters. No memory of the initial state (for a
given mean energy) is thus found on simple observables. These results
agree well with the previous findings of Ref.~\onlinecite{Manmana2007}
on a similar model. Having in mind this qualitative argument, the
results of Ref.~\onlinecite{Kollath2007} on the non-integrable
one-dimensional (1D) Bose-Hubbard model (BHM) look rather
counter-intuitive: for small quenches, a thermalized regime was found
in the sense that two independent observables computed within a
(grand)-canonical ensemble (and not microcanonical) and from
time-evolution gave the same results. On the contrary, a mean-field
treatment of the 1D BHM interpreted in the framework of chaos
theory~\cite{Cassidy2009} supports non-thermalization below an
interaction threshold and thermalization above (mean-field theory is
however known to fail for this strongly-correlated model so the
results are not under control). The findings of
Ref.~\onlinecite{Kollath2007} were later supported by the calculation
of the diagonal ensemble distributions which looked like an
\textit{approximate} Boltzmann law~\cite{Roux2009} in the small quench
regime. Surprisingly, for large quenches, a non-thermalized regime was
found in Ref.~\onlinecite{Kollath2007} in which the correlations bear
a strong memory of the initial state (in the sense that they are
closer to the ones in the initial state than to the thermalized
ones). This non-equilibrium behavior was attributed to
the very peculiar shape of the diagonal ensemble in this
regime~\cite{Roux2009}. An important step towards the understanding of
the non-thermalized regime on finite size systems was made very
recently~\cite{Biroli2009} by giving numerical evidences on the 1D BHM
that ETH does not apply for large quenches in finite systems and suggesting a general
framework in terms of rare events contributing to the distribution, providing 
a refined version of the ETH.

As integrability is often one of the ingredients that play a role in
the physics of quenches, we briefly recall that, for 1D quantum
many-body models, integrability can be well defined for a class of models which have the property of
scattering without diffraction~\cite{Sutherland2004}. This has two
consequences that are in relation with the question of thermalization:
the momenta of the particles do not redistribute~\cite{Sutherland2004} (a process which is
believed to be essential to get the thermalized momentum
distribution), and there is an extensive number of conserved
quantities that separate the eigenstates in many sectors, constraining
the time evolution. In the context of nuclear physics, random-matrix
theory has been proposed to describe the statistical features of the
bulk of the spectrum and it is commonly conjectured that
non-integrable quantum many-body or classically chaotic models display
universal level statistics~\cite{RMT}. Level
statistics have been computed in a few many-body
models~\cite{RMT-Many-body}, supporting the conjecture, but these
results are restricted to a few models and it cannot be excluded that
diffractive models could display non-universal level statistics. The
Bose-Hubbard model is a bit peculiar in this sense: if one denotes by
$N_{\textrm{max}}$ the maximum number of bosons onsite, the model is
non-diffractive only for $N_{\textrm{max}}=1$~\cite{Kollath2010}. In
addition, if $U$ is the interaction strength, $U=0$ is an integrable
point (the atomic limit $J=0$ is as well exactly solvable). Level statistics and delocalization properties of the
eigenstates have shown~\cite{Kollath2010, Kolovsky2004} that the BHM
display features of quantum chaotic systems for non-zero $U$ (and
larger $N_{\textrm{max}}$).

The first goal of this paper is to discuss the crossover from small to
large quench amplitude regimes on the basis of energetic and static
fidelity arguments, and to evaluate the finite size effects that are
associated to this crossover. We then turn to a detailed discussion of
the diagonal ensemble and the verification of the ETH in the BHM,
complementary to what has been done in Refs.~\onlinecite{Roux2009} and
\onlinecite{Biroli2009}. We show that the observed Boltzmann-like
regime is spoiled by strong finite size effects that prevent both an
accurate definition of an effective temperature and the comparison
with the microcanonical ensemble. In the large quench limit, we
explain in details that the breakdown of the ETH is actually related to
the ``integrable'' quench limit $U_i=0 \rightarrow U_f=\infty$. Thus,
non-thermalization in the 1D BHM is, on finite systems, reminiscent of
the atomic limit. While the $U=0$ limit of the Bose-Hubbard model is
trivially integrable as a free boson model, the infinite $U$ (or
atomic) limit is a bit particular: for very large $U$ \emph{and}
focusing on the \emph{low-energy} part of the spectrum, the model is
effectively identical to an integrable 1D hard-core bosons model
($N_{\textrm{max}}=1$). However, we will see that, to understand the
large-$U$ limit of the quench, we will have to consider the whole
excitation spectrum and not only the low-energy part. This result can
be qualitatively and partially connected to the effect of the
proximity to integrable points in quantum quenches, studied very
recently in fermionic and hard-core bosonic models~\cite{Rigol2009,
  Rigol2009a}, in the sense that the observed non-thermalized regime
\emph{on finite systems} is connected to a particular limit in which
the model has high degeneracies. Throughout the paper, we also give a
simple but interesting example of a quench in a toy model consisting
of free bosons confined in an harmonic trap. The motivation for it is
that it surprisingly shares some qualitative features with the 1D BHM
and that it allows for analytical calculations on some properties of
the diagonal ensemble distribution. This model also corresponds to a
standard experimental setup (so for the BHM) although interactions
would have to be taken into account for a realistic comparison.

The paper is organized as follows: we first review in
Sec.~\ref{sec:model} the definitions of the time-averaged
density-matrix, the ETH and the computation of the diagonal weights
for the two models under study. In Sec.~\ref{sec:finitesize}, we
suggest two kinds of crossover number of particles to distinguish the
small and large quench regimes. Lastly, we discuss in
Sec.~\ref{sec:thermalization} the fate of the ETH in the 1D BHM and on
small finite size systems.

\section{Models and computation of the weights of the diagonal ensemble}
\label{sec:model}

\subsection{The time-averaged density-matrix and the ``eigenstate
  thermalization hypothesis'' }

As discussed in recent papers~\cite{Rigol2008, Roux2009,
  Faribault2009, Rigol2009, Rigol2009a, Biroli2009}, the time-averaged
expectation values of any observable are governed by the time-averaged
density-matrix $\bar{\rho}$, which is diagonal in the final
Hamiltonian eigenstate basis, provided the spectrum is
non-degenerate. From now on, we only consider finite size systems that
have a discrete spectrum. This leads to the so-called ``diagonal
ensemble'' that has weights fully determined by the overlaps between
the initial state $\ket{\psi_{0,i}}$ and eigenstates
$\ket{\psi_{n,f}}$ of the final Hamiltonian $\Ham_f$. Usually,
$\ket{\psi_{0,i}}$ is the ground-state of the initial Hamiltonian
$\Ham_i$ and we assume in the following that we start from this
zero-temperature pure state. We also consider that the final
Hamiltonian takes the form
\begin{equation}
\label{eq:general_Ham}
\Ham_f = \Ham_i + \lambda \Ham_1\;,
\end{equation}
where $\lambda$ (that has the dimension of an energy) is called the
quench amplitude, and $\Ham_1$ is the dimensionless perturbing
operator. Working on a \emph{global} quantum quench means that
$\Ham_1$ is assumed to be an extensive operator that scales with the
number of particles $N$. The time-averaged density-matrix is defined
by $\bar{\rho} = \lim_{t \rightarrow \infty} \frac 1 t \int_0^{t}
\ket{\psi(s)}\bra{\psi(s)}ds$ with $\ket{\psi(t)} = e^{-i\Ham_f t}
\ket{\psi_{0,i}}$. It is important to realize that the infinite time
limit is taken before the thermodynamical limit. If the spectrum has
exact degeneracies, the time-averaged density-matrix reads:
\begin{equation}
\label{eq:tadm}
  \bar{\rho} = \sum_n p_n \ket{\psi_{n,f}}\bra{\psi_{n,f}} + \sum_d \ket{\psi_{d,f}}\bra{\psi_{d,f}}
\end{equation}
where $n$ labels non-degenerate eigenstates of $\Ham_f$ and $p_n =
\md{\inter{\psi_{n,f}}{\psi_{0,i}}}^2$ are the diagonal weights. $d$
labels the basis of the degenerate subspaces, and the vectors $
\ket{\psi_{d,f}} = \sum_{q_{d,f}} \inter{q_{d,f}}{\psi_{0,i}}
\ket{q_{d,f}} $ keep a memory of the initial phases of
$\ket{\psi_{0,i}}$ with respect to the $\ket{q_{d,f}}$.  In the
situation where $\bar{\rho}$ is block-diagonal, in order to get
time-averaged results for an observable $\mathcal{O}$ which has
off-diagonal matrix elements in the $\Ham_f$ eigenstate basis, one
would have to compute all the overlaps $\inter{q_{d,f}}{\psi_{0,i}}$
and $\elem{q'_{d,f}}{\mathcal{O}}{q_{d,f}}$ and sum up the
contributions of all a degenerate subspace. In the following, this
would be the case only for the free boson model and we will actually
only use observables that are diagonal because the dimensions of the
degenerate sectors grows (roughly) exponentially with the number of bosons
$N$. For the Bose-Hubbard model, one can check that the spectra are
non-degenerate in each symmetry sector.

For a generic non-integrable model, the ``eigenstate thermalization
hypothesis'' (ETH) has been surmised~\cite{Deutsch1991, Srednicki1994,
  Srednicki1994a, Srednicki1999, Rigol2008}, suggesting an explanation
for thermalization in an isolated quantum system and a justification
for the use of the microcanonical ensemble. The ETH is supported by
semi-classical and random-matrix theory arguments~\cite{Peres1984, Deutsch1991,
  Srednicki1994, Srednicki1994a, Srednicki1999}, and was checked
numerically on particular models~\cite{EarlyETH, Rigol2008, Rigol2009,
  Rigol2009a}. The ETH boils down to the fact that, in a given
small window of energy, the diagonal observables $\mathcal{O}_n =
\elem{\psi_{n,f}}{\mathcal{O}}{\psi_{n,f}}$ that contribute to the
time-averaged expectation value $\bar{\mathcal{O}} =
\textrm{Tr}[\bar{\rho} \mathcal{O} ] = \sum_{n} p_n \mathcal{O}_n$
hardly depend on the eigenstate $n$ (in short, 
$\mathcal{O}_n \simeq \bar{\mathcal{O}}$ in a small energy
window). Consequently, any distribution peaked around the mean energy,
and one can show on general grounds that the relative width of the
distribution scales to zero as $N^{-1/2}$~\cite{Rigol2008} (although
some slower scalings could occur~\cite{Roux2009}), will give the same
observables as the microcanonical ensemble, therefore accounting for
thermalization. For integrable models~\cite{Rigol2008, Rigol2009,
  Rigol2009a}, non-thermalization is explained from the fact that
observables fluctuate a lot within a given energy window, which may be
associated with the extensive number of conserved quantities that
exist in these models. A more subtle scenario for the breakdown of the
ETH was recently proposed~\cite{Biroli2009}, in which some ``rare''
states have a significant contribution to the averaged observables.

\subsection{Free bosons in an harmonic trap}

We now describe how to get the diagonal weights for two particular
models. Firstly, we consider a model of $N$ non-interacting bosons
initially confined in an harmonic trap of frequency $\omega_i$ and
lying in the zero-temperature ground-state. The frequency is changed
to $\omega_f$ at time $t=0$. For this model, the quench amplitude is
defined as $\lambda = \omega_f/\omega_i - 1$ (taking $\omega_i$ as the
unit of energy), according to the expression of the quench parameter
in terms of the harmonic oscillator ladder operators.  We start with
the computation of the single-particle wave-function overlaps
$\mathfrak{p}_n$ since the results for the many-body wave-function will
be expressed as a function of them. The single-particle spectrum is
non-degenerate and the single-particle eigenfunctions are:
\begin{equation*}
\phi_n(x) = \frac{1}{\sqrt{2^n n!\sqrt{\pi}\sigma}} e^{-\frac{x^2}{2\sigma^2}} H_n\left(\frac{x}{\sigma}\right)\;,
\end{equation*}
with $\sigma = \sqrt{\hbar/m\omega}$ and $H_n$ the Hermite
polynomials. The single-particle excitation spectrum is split into the
odd and even parity sectors and the overlaps are non-zero for
even-parity wave-functions only. They read:
\begin{equation}
\label{eq:single-particle-weights}
\mathfrak{p}_{2n} = \frac{(2n)!}{2^{2n}(n!)^2} \frac{\sqrt{1+\lambda}}{1+\lambda/2}\left(\frac{\lambda}{\lambda+2}\right)^{2n}
\end{equation}
for integer $n$. The many-body wave-function of an $N$-bosons excited
configuration $\{n_j\} = \{n_0,\cdots,n_{m}\}$ of the final
Hamiltonian $\Ham_f$ (with highest occupied level $m$) is:
\begin{equation*}
  \ket{\{n_j\}} = \sqrt{\frac{n_0!n_1!\cdots n_m!}{N!}} 
  \sum_{p \in \mathcal{P}} \ket{\phi_{1,f}{:}{p(1)},\cdots,\phi_{m,f}{:}{p(N)}}\;.
\end{equation*}
with $\mathcal{P}$ the set of all permutations and $n_j$ the
occupation of the single-particle orbital $\phi_{j,f}$. Overlapping
this state with the $N$-bosons initial ground-state
$\ket{\phi_{0,i},\cdots,\phi_{0,i}}$ gives the many-body weights
\begin{equation}
\label{eq:many-particle-weights}
  p_{\{n_j\}} = N! \frac{(\mathfrak{p}_0)^{n_0}}{n_0!} 
  \frac{(\mathfrak{p}_2)^{n_2}}{n_2!}\cdots \frac{(\mathfrak{p}_{m})^{n_{m}}}{n_{m}!}
\end{equation}
In this equation, all $m$'s are even integers. The total energy of this
excitation is $E_{\{n_j\}} = \hbar \omega_f (2 n_2 + 4n_4 + \cdots + m
n_{m} ) + \hbar \omega_f N/2$ with the constraint $\sum_{j=0}^{m/2}
n_{2j}=N$. Eq.~(\ref{eq:many-particle-weights}) is nothing but the
multinomial distribution associated with the elementary
probabilities $\mathfrak{p}_m$ and it is thus clear that it is
normalized. We also see that formula~(\ref{eq:many-particle-weights})
is in general valid for a free boson model, starting from the
condensed ground-state (and specifying the $\mathfrak{p}_m$). If one
takes the single-particle Boltzmann factor for the $\mathfrak{p}_m$,
one recovers the many-body Boltzmann factor for the
configuration. Contrary to statistical ensemble distributions, the
weights do not show a simple dependence of the configuration energy. This quench is qualitatively similar to a Joule
compression/expansion as the 1D effective density $n=N\omega$ suddenly
changes. In fact, $\lambda=n_f/n_i-1$ is related the ratio of the
effective densities. Other examples of quantum mechanical treatments
of the Joule expansion can be found in the
literature~\cite{Camalet2008, Aslangul2008}.

In order to get the distribution of the weights versus energy, we
resort to numerics: using a fixed number of low-lying \textit{even
parity} levels $N_s$, we scan all possible configurations of $N$
bosons in these $N_s$ levels iteratively up to roughly $62\times 10^9$
configurations ($N=18$ and $N_s=22$). The truncation error associated
with a finite $N_s$ is checked by summing up the weights.

\subsection{The one-dimensional Bose-Hubbard model}

The Bose-Hubbard model in a one-dimensional lattice, known to be
non-integrable for $U \neq 0$, is described by the following
Hamiltonian:
\begin{equation*}
\Ham = -J \sum_j [b^{\dag}_{j+1} b_j + b^{\dag}_{j} b_{j+1} ] 
        + \frac{U}{2} \sum_j n_j(n_j-1)\,,
\end{equation*}
with $b^{\dag}_j$ the operator creating a boson at site $j$ and $n_j =
b^{\dag}_j b_j$ the local density. $J$ is the kinetic energy scale
while $U$ is the magnitude of the onsite repulsion. In an optical
lattice, the ratio $U/J$ can be tuned by changing the depth of the
lattice and using Feshbach resonance. When the density of bosons is
fixed at $n=1$ and $U$ is increased, the zero-temperature equilibrium
phase diagram of the model displays a quantum phase transition from a
superfluid phase to a Mott insulating phase in which particles are
localized on each site. The critical point has been located at $U_c
\simeq 3.3 J$ using numerics~\cite{Kuhner2000}. The quenches are
performed by changing the interaction parameter $U_i \rightarrow U_f$
(we set $J=1$ as the unit of energy in the following), so we have
$\lambda = (U_f - U_i)/2$, and the perturbing operator $\Ham_1 =
\sum_j n_j (n_j-1)$ is diagonal.  Numerically, one must fix a maximum
onsite occupancy $N_{\textrm{max}}$ and we take $N_{\textrm{max}}=4$
unless stated otherwise. Exact diagonalization calculations are
carried out using periodic boundary conditions and translational
invariance. We denote by $0 \leq k \leq L-1$ the total momentum
symmetry sectors. The algorithm to get the ground-state and
eigenstates of the Hamiltonian is a full diagonalization scheme for
sizes up to $L=10$ at unitary filling. For some of the quantities, we
use the Lanczos algorithm up to $L=15$. In Ref.~\onlinecite{Roux2009},
the Lanczos algorithm has been proposed to compute the low-energy
weights of the distribution. This worked relatively well for the 1D
BHM, and in particular for the spectrum-integrated quantities but it
may not be suited for all possible kind of quenches. We notice that in
the case of quenches with a mean energy deep in the bulk of the
spectrum, a generalization of the Lanczos
algorithm~\cite{Ericsson1980} that works in the bulk of a spectrum
could be used to get the main weights. For what we call small quenches
in the following, the larger weights are in the low-energy region so
Lanczos can generically give good results in such situations.

\section{Arguments on finite size effects and the different regimes of
  a quantum quench}
\label{sec:finitesize}

The goal of this part is to quantify the distance of the quench
distribution $\bar{\rho}(E)$ from the many-body ground-state and the low-energy region of the
spectrum. A first distance is defined from an energetic argument and a
second one from the overlap with the ground-state of $\Ham_f$. Both
criteria leads to a crossover number of bosons $N_c(\lambda)$ that can
be computed numerically and that diverge with small $\lambda$ as a
power-law. When $N \ll N_c$, the quench probes the low-energy part of
the spectrum while when $N \gg N_c$, high-energy physics govern the
time-evolution. Both definitions do not depend on the integrability of
the model but we may argue that for non-integrable models, there is a
strong qualitative difference between the low-energy part of the
spectrum and the bulk of the spectrum. These finite-size effects are
rather generic while other kind of finite-size effects can emerge for
a given model: this will be for instance the case for the BHM at
large $U$.

\subsection{Crossover number of particles from an energetic argument}
\label{sec:energetic}

\begin{figure}[b]
\centering
\includegraphics[width=.85\columnwidth,clip]{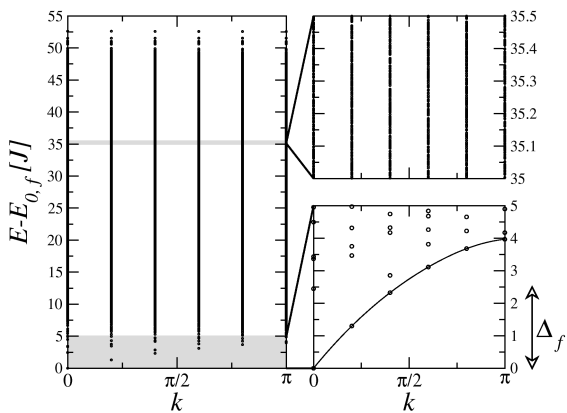}
\caption{A typical many-body spectrum of a finite-size system: this
  example is taken from the 1D Bose-Hubbard model with $U_f/J=2.5$ and
  $L=N=10$. Energies are given as a function of the total momentum
  $k$. The width of the spectrum is typically proportional to $N$ or
  $N^2$ depending on the statistics. Zooms in the low-energy region
  and in the bulk of the spectrum (grey region) are given. The
  low-energy region features elementary excitations up to a typical
  energy scale $\Delta_f$ which is assumed to be microscopic,
  i.e. non-extensive. Here, we take $\Delta_f=U_f$ and the relation
  dispersion of the excitation branch is sketch (the line is a guide
  to the eyes).}
\label{fig:generic_spectrum}
\end{figure}

\emph{The low-energy part of the spectrum} -- We first have to specify
what we mean by the low-energy region of the spectrum: it corresponds
to the typical energies of a few elementary excitations above the
ground-state. These elementary excitations are quasi-particles,
collectives modes, particle-hole excitations\dots Single or few
excitations give a structure (dispersion relations, continuum of
low-lying excitations) to the low-energy part of the many-body spectrum (see an
example in Fig.~\ref{fig:generic_spectrum}). We denote by
$\Delta_f$ the typical energy scale of a single excitation, it is
a \emph{microscopic} energy scale. In Bethe-ansatz solvable or free systems, a high energy
excitation can be understood as a superposition of single-particle excitations
while this is no longer true for non-integrable systems~\cite{RMT,
  RMT-Many-body}. If the number of elementary excitations remains
small enough, they may hardly interact and have integrable-like
features in the low-energy part of the spectrum. We thus expect a
smooth crossover between integrable-like and non-integrable like
behaviors with increasing the energy above the ground-state, but
the typical energy of this crossover is hard to evaluate, except that
it must be above $E_{0,f} + \Delta_f$.

\begin{figure}[t]
\centering
\includegraphics[width=.85\columnwidth,clip]{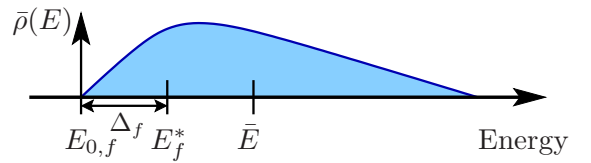}
\caption{(Color online) Sketch of the energy scales in a quantum
  quench. The initial state builds up an energy distribution
  $\bar{\rho}(E)$ (diagonal ensemble) around a mean energy $\bar{E}$
  fixed by the initial state. The quench amplitude $\lambda$ tunes
  both $\bar{E}$ and the ground-state energy $E_{0,f}$ and the
  low-energy scale $E_f^*$.  $\Delta_f = E_f^* - E_{0,f}$ is assumed
  to be non-extensive while $E_{0,f}$ and $\bar{E}$ are
  extensive. $\bar{E} = E_f^*$ defines the crossover number of
  particles $N_c$. In the thermodynamical limit, one expects $\bar{E}
  \gg E_f^*$ for any finite $\lambda$.}
\label{fig:sketch}
\end{figure}

\emph{Criteria} -- We consider that the energy
distribution $\bar{\rho}(E)$ is centered around the the mean energy $\bar{E} =
\elem{\psi_{0,i}}{\Ham_f}{\psi_{0,i}}$ of the distribution (fixed by
the initial state) as in general $\Delta E/(\bar{E}-E_{0,f}) \sim
1/\sqrt{N}$. Since $\ket{\psi_{0,i}}$ is not an eigenstate of
$\Ham_f$, we necessarily have $\bar{E} > E_{0,f}$. The criteria we
choose to distinguish between low-energy (or small) quenches and
high-energy (or large) quenches is $\bar{E} = E^*_f$ (see
Fig.~\ref{fig:sketch}) where $E_f^*$ is such that $E_f^*-E_{0,f} =
\Delta_f$ with the ground-state energy $E_{0,f}$.  It corresponds to
the situation where the mean-energy put into the system excites
roughly only one elementary excitation and is thus a finite-size
effect. Another way to introduce the same criteria is the following:
$(\bar{E}-E_{0,f})/\Delta_f$ is the energy difference between the
initial state and the final ground-state in units of the typical
elementary excitation energy $\Delta_f$, the criteria corresponding to
a distance of one $\Delta_f$~\footnote{We notice that $\Delta_f$ is
  different from the finite size gap to the first excitation (there
  can be a huge number of states between $E_{0,f}$ and
  $E_f^*$).}. The criteria thus amounts to a lower bond of the energies
  at which one enters in the bulk of the spectrum. 
  The order of magnitude of $\Delta_f$ is set by the microscopic
units of energy of the model. For instance, we'll take $U_f$ in the
BHM as it controls the sound velocity in the superfluid region and the
Mott gap in the Mott phase.

This criteria gives a relation between the crossover number of
particles $N_c$ (on a lattice we work at finite density so it also
corresponds to a crossover length $L_c$) and the quench amplitude $\lambda$
is such that, if $N \ll N_c(\lambda)$, the energy is mostly distributed among the
low-energy excitations while if $N \gg N_c(\lambda)$, most of the
weights are on high-energy excitations. We can rewrite the criteria in
a more tractable way: using the notation $e=E/N$ for the energy per
particle, and the label $0$ for ground-state energies, it reads
\begin{equation}
\label{eq:N_c_from_energy}
 N_c(\lambda) = \frac{\Delta_f(\lambda)}{\bar{e}(\lambda) - e_{0,f}(\lambda)}\;.
\end{equation}
Interestingly, we expect $N_c(\lambda)$ to generically diverge as
$\lambda^{-2}$ in the limit of small $\lambda$. Indeed, we have
$\bar{e} = e_{0,i} + \lambda h_{1,i}$ with $h_{1,i} =
\elem{\psi_{0,i}}{\Ham_1}{\psi_{0,i}}/N$, the expansion $e_{0,f}
\simeq e_{0,i} + (de_0/d\lambda)_i\lambda + (d^2e_0/d\lambda^2)_i
\lambda^2/2$, and $(de_0/d\lambda)_i = h_1$ after Feynman-Hellman
theorem. With Eq.~(\ref{eq:general_Ham}), one finally gets $N_c(\lambda)
\lambda^2 \rightarrow 2 \Delta_i / (d^2e_0/d\lambda^2)_i$ at $\lambda \rightarrow 0$.

A few comments can be made on the criteria:
\begin{itemize}
\item When comparing quenches from the same initial state but with
  different $\Ham_f$, $\lambda$ controls the mean energy per particle
  put into the system. Thus, $\lambda$ is a meaningful parameter even
  in the thermodynamical limit.
\item This definition looks qualitative due to the rather arbitrary
  choice of $\Delta_f$ and to the fact that, on finite systems, the
  energy distribution can have a rather large width associated with
  energy fluctuations $\Delta E$. We point out that $N_c$ is a
  crossover number so that $N \simeq N_c$ has no particular
  meaning. Furthermore, from the divergence at small $\lambda$, one
  can have $1 \ll N \ll N_c$, i.e. a situation where energy
  fluctuations vanish.
\item When $\lambda$ is scanned from $0$ to a finite value, both the
  mean energy and the region of the spectrum that plays a role in the
  time-evolution (around $\bar{e}$) are continuously changed. One can
  also notice that a quench that starts from a ground-state does not
  necessarily allow to access any energy of the $\Ham_f$ spectrum, 
  contrary to the situation where one prepares the initial state at
  will.
\item The regimes $N \gg N_c$ and $N \ll N_c$ are expected to be
  physically different for generic (non-integrable) systems. Below
  $\Delta_f$, the density of states is usually much smaller than in
  the bulk of the spectrum: level spacings are of order of $1/N$ and
  observables can strongly fluctuate with the eigenstate number as it
  can be seen in Figs.~\ref{fig:observables_g1} and
  \ref{fig:observables_nk0} (similar observations can be made in the
  figures of Refs.~\onlinecite{Rigol2008, Rigol2009, Rigol2009a}). In
  this low-energy region, RMT arguments are not expected to
  work~\cite{RMT} and the eigenstates may not be ``typical'' so we
  expect the ETH to fail. These qualitative observations support the 
  difference between the low-energy region and the high-energy region
  of the spectrum made at the beginning of this section.

  As the full spectrum width grows as $N$ or $N^2$ (depending on the
  statistics of the particles) while the number of eigenstates grows
  exponentially with $N$, the density of states in the bulk of the
  spectrum is exponentially large. In this ``high-energy'' regime
  (with respect to elementary excitations), semi-classical and RMT
  arguments are believed to work reasonably well for
  \emph{non-integrable} models~\cite{RMT}, which was checked on some
  strongly correlated systems~\cite{RMT-Many-body}. As observed
  numerically on several examples~\cite{Rigol2008, Rigol2009,
    Rigol2009a}, simple observables hardly depend on the eigenstate
  number in this regime, supporting the ETH.
\item In the thermodynamical limit, we always have $N \gg N_c$ and the
  small quench regime is thus expected to vanish. If one wants to
  check the ETH on a finite size systems, one needs sufficiently large
  $\lambda$ in order to try to reach the bulk of the
  spectrum. However, we will see in this paper a counter-example (the
  BHM) where ETH fails at large $\lambda$ (see also Ref.~\onlinecite{Biroli2009}). 
  Even though it looks
  difficult to use quenches to probe very low-energy excitations in a very
  large system, on a finite system, one could tune the mean energy
  from the low to high energy part of the spectrum using
  $\lambda$. Furthermore, this small quench regime is certainly of
  interest for numerical simulations, and also for experiments using a
  relatively small number of atoms (few hundreds or thousands).
\item Lastly, it could be interesting to compare this criteria with
  the domain of validity of bosonization~\cite{Cazalilla2006,
    Iucci2009} and conformal field theory~\cite{Calabrese2006,
    Calabrese2007} but this is beyond the scope of this paper. We note
  that conformal field theory can describe accurately quenches in
  certain integrable models in the thermodynamical limit and for
  arbitrary quench amplitudes~\cite{Calabrese2006, Calabrese2007}. 
  Non-integrable models low-energy features that are described in
  terms of a free particles (integrable) theory, as bosonization, should
  display non-thermalized features as for integrable models. In this
  respect, Ref.~\onlinecite{Barmettler2009} gives interesting examples
  on the applicability of these methods to the quench situation.
\end{itemize}

We now give examples of $N_c(\lambda)$ for the two models under study.
In the free boson model, the mean energy after the quench can be
computed analytically:
\begin{equation*}
  \bar{e} = e_{0,i} +  \frac{\hbar\omega_f}{4}
  \left(\frac{\omega_f}{\omega_i} - \frac{\omega_i}{\omega_f}\right)\;,
\end{equation*}
with $e_{0,i/f} = \hbar \omega_{i/f}/2$. The energy fluctuations are
given by $\Delta e = (\bar{e} - e_{0,i})\sqrt{2/N}$, showing that the
distribution gets peaked in the thermodynamical limit with the usual
scaling. A natural choice for $\Delta_f$ is $\hbar\omega_f$ (the only
microscopic energy scale) and the crossover number of bosons can be
expressed as a function of the quench amplitude:
\begin{equation*}
N_c = \frac{\hbar \omega_f}{\bar{e} - e_{0,f}}
= 4 \left(\frac{\omega_i}{\omega_f} + \frac{\omega_f}{\omega_i}-2 \right)^{-1}
=  4 \frac{\lambda + 1}{\lambda^2}\;.
\end{equation*}
This expression diverges as $4/\lambda^2$ in the small quench regime
and vanishes as $4/\lambda$ in the large quench regime.

\begin{figure}[t]
\centering
\includegraphics[width=.85\columnwidth,clip]{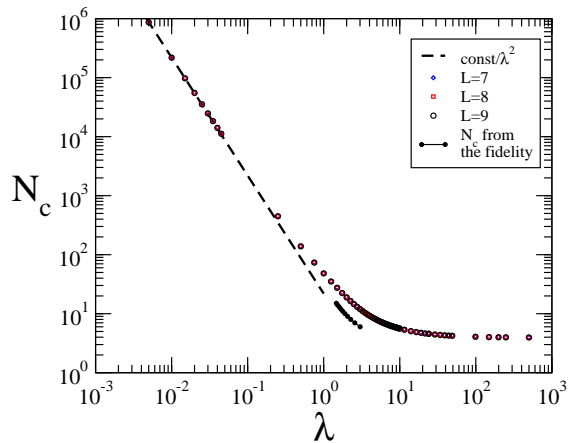}
\caption{(Color online) Crossover number of bosons $N_c$ obtained from
  the energetic argument in the 1D BHM at filling $n=1$ and starting
  from $U_i=2$. A few points obtained from the criteria based on the
  static fidelity are also given.}
\label{fig:BHM_Nc}
\end{figure}

For the 1D BHM, we take $\Delta_f = U_f$ and $N_c$ is given in
Fig.~\ref{fig:BHM_Nc} for the particular initial value $U_i = 2$. It
displays the expected $\lambda^{-2}$ divergence at small quenches.  We
notice that the finite size effects on this energy-based criteria are
pretty small. This can be put on general grounds for 1D systems: for
critical systems the finite-size effects on the ground-state energy
per particles have a universal correction~\cite{Affleck1986}:
\begin{equation*}
e_0(L) = e_0(\infty) - \frac{c\pi u}{6L^2} + O\left(\frac{1}{L^2}\right) 
\end{equation*}
with $u$ the sound velocity and $c$ the central charge. If the system
is gapped, the corrections are even smaller as they are exponentially
suppressed, by a factor $\exp(-L/\xi)$ with $\xi$ the correlation length
enters in the formula. In the large quench limit of the BHM, one can
argue that $N_c$ saturates to a finite value. Indeed, in the limit of
large $\lambda$, one finds that $N_c \rightarrow
2/(\moy{n^2}_{0,i}-\moy{n^2}_{0,f}) + \mathcal{O}(1/\lambda) \simeq
2/\moy{n^2}_{0,i}$, as the density fluctuations $\moy{n^2}_{0,f}$ are
suppressed in the Mott phase.  Notice that the energy fluctuations,
that scale as $N^{-1/2}$ in the 1D BHM, have been computed numerically
in Ref.~\onlinecite{Roux2009}. The full curve and the two asymptotic
behaviors can be simply computed from ground-state calculations.

\subsection{Crossover number of particles based on the static fidelity}
\label{sec:fidelity}

In the thermodynamical limit, the (squared) fidelity between the two
ground-states $\mathcal{F} =
\vert{\inter{\psi_{0,i}}{\psi_{0,f}}}\vert^2$ is generally expected to
vanish exponentially with the system size or number of
particles. Interestingly, $1-\mathcal{F}$ counts the contribution of
the excited states to the time-evolution. A possible definition of a
crossover number of particles can thus be the value of $\lambda$ and
$N$ such that $\mathcal{F} = 1/2$, i.e. half of the total weight in
the ground-state and half in the excited states. In the limit $\lambda
\rightarrow 0$, one can introduce the fidelity susceptibility
$\chi_{i,L}$ through the expansion $\mathcal{F} \simeq 1 - \lambda^2
\chi_{i,L}/2$. The scaling of $\chi_{i,L}$ is in general
non-trivial. If $\Ham_i$ is gapped, the scaling $\chi_{i,L} \sim L$
has been proposed~\cite{Venuti2007}, which gives the divergence $N_c
\sim \lambda^{-2}$. In critical systems, super-extensivity,
corresponding to a scaling $\chi_{i,L}/L \sim L^{\alpha_i}$ with
$\alpha_i >0$, can occur~\cite{Venuti2007, Cozzini2007}, leading to a
slower divergence $N_c \sim \lambda^{-2/(1+\alpha_i)}$ that depends on
the initial state. Notice that we qualitatively expect that the $N_c$
from the fidelity will be smaller than the one based on energetic
argument because, on sufficiently large systems, $\mathcal{F}$ can be
very small while the mean energy is still in the low-energy part of
the spectrum.

For the free boson model, the static fidelity as a function of
$\lambda$ is $\mathcal{F} = (\sqrt{1+\lambda} /
(1+\lambda/2))^N$. Setting $\mathcal{F} = 1/2$, one has the crossover
number of bosons $N_c$:
\begin{equation}
N_c = \frac{\ln 2}{\ln \left( \frac{1+\lambda/2}{\sqrt{1+\lambda}} \right)}
\end{equation}
Notice that it also diverges in the small quench regime as $N_c = 8
\ln 2/ \lambda^2$ with the same power-law as for the energetic
arguments. Put in other words, this means that the \emph{many-body} ground-state
occupation is robust within a $25\%$ change in $\omega$ for $N=10^2$,
$7\%$ for $N=10^3$ and $2\%$ for $N=10^4$ (see next section for the
\emph{single-particle} level occupation). In the large amplitude
limit, it decreases only logarithmically with $\lambda$, $N_c \simeq 2
\ln 2 /\ln \lambda$ but the prefactor is already small.

The fidelity can also be computed for the 1D BHM by Lanczos
calculations. Using the curves $\mathcal{F}(\lambda)$ obtained
numerically, we determined $N_c(\lambda)$ for numbers of bosons from 6
to 15. The result is plotted in Fig.~\ref{fig:BHM_Nc}. Due to the
relatively small sizes accessible with Lanczos, we cannot investigate the scaling
exponent of the small quench divergence. The ground-state fidelity of
the 1D BHM has been studied in Ref.~\onlinecite{Buonsante2007}.  We
observe that the static fidelity could be computed on larger chains
with matrix-product
states based algorithms~\cite{Vidal2004, McCulloch2007} or quantum
Monte-Carlo techniques~\cite{Schwandt2009}.

\subsection{Quench and transition temperature to the Bose-condensed
  regime in the free bosons model}

The free bosons model undergoes a transition to a Bose-condensed state
below a critical temperature $T_c$. In the 1D harmonic trap and on a
finite size system, the lowest single particle level occupation
$\moy{n_0}$ becomes of the order of $N$ below $T_c \simeq \hbar \omega
N / \ln(N)$ (standard calculations of $T_c$ are performed in the
grand-canonical ensemble and one sees that for fixed effective density $\omega
N$ and $N\rightarrow \infty$, $T_c \rightarrow 0$ in agreement with
the fact that their is no Bose-condensation in this model in the
thermodynamical limit although condensed and non-condensed regimes are
clearly seen on finite systems). This critical temperature corresponds to a
critical energy $E_c - E_0 \sim  \hbar \omega N^2$. These standard results can be used to answer the
question : whether or not a large quench from the many-body
ground-state can drive the system into the non-condensed regime? We
found that the mean-energy put into the system scales as $\bar{E} \sim
E_{0,f} + \hbar \omega_f N \lambda$ so that $\lambda \sim N$ is
required to reach $E_c$ and the non-condensed regime. This surprising behavior
(diverging with the number of bosons) actually agrees with the
exact scaling of the single-particle ground-state occupation number
which can be computed for the quench since we have seen that the
distribution is the multinomial one : we have $\moy{n_0} = N p_0 \sim
N/\sqrt{\lambda}$ at large $\lambda$. Similarly, the fluctuations can
be computed and read $\moy{n_0^2-\moy{n_0}^2} = Np_0(1-p_0)$ so that
the relative fluctuations scale as $1/\sqrt{N}$ with a
$\lambda$-dependent prefactor. Consequently, starting from the
many-body ground-state (for which $\moy{n_0} = N$), one stays in the
condensed regime for finite $\lambda$ and one needs $\lambda \sim
N^{z}$ with $z>2$ to make $\moy{n_0}$ scale to zero in the
thermodynamical limit. The physical origin of the fact that the quench
process makes it difficult to reach the critical temperature is that
the many-body ground-state has vanishing overlaps with the excited states
above $T_c$ because they have negligible contributions from the single-particle
ground-state. Starting from a finite temperature state, the quench
could help cross the critical temperature.

\section{Diagonal ensemble and thermalization}
\label{sec:thermalization}

In this section, we compare averages of the expectation values of
observables obtained from different ensembles: the diagonal,
microcanonical and canonical ones. We also show the behavior of some
local and global observables as a function of the energy per particle
to discuss the possibility of thermalization according to the ETH.
The first numerical evidences that the ETH does not work for large
quenches on finite systems of the 1D BHM were recently given in
Ref.~\onlinecite{Biroli2009}.

\subsection{Microcanonical temperature and the density of states}

As a preliminary, we discuss the finite size effects and possible
issues with the microcanonical ensemble in the model under study. The
standard way to define the microcanonical temperature $T_M$ of a
closed system is from Boltzmann's formula
\begin{equation}
\frac{1}{T_M} = \frac{\partial s_M}{\partial \bar{e}}\;,
\end{equation}
where we use the entropy per particle $s_M=S_M/N$ and the statistical
entropy $S_M(\bar{E}) = k_b \ln \Omega(\bar{E})$. $\Omega(\bar{E})$ is
the number of states within a small energy window $\delta E$ around
$\bar{E}$. Any distribution that is peaked enough ($\delta E/\bar{E}
\rightarrow 0$ in the thermodynamical limit) will pick up the local
density of states $g(\bar{e})$ through $\Omega(\bar{E}) \simeq
g(\bar{E}) \delta E$. Usually, $\delta E$ is taken as the energy
fluctuations with $\delta E \sim \bar{E}/\sqrt{N}$. Thus, $\delta E$
is typically much larger than microscopic energy scales such as
$\Delta_f$. For the free boson model, energies per particle are
separated by $\hbar \omega_f/N$ and the degeneracy $g(e)$ of each
level can be computed numerically for small systems. Asymptotic
analytical results exist in the large energy limit for
$g(e)$~\cite{Abramowitz1965, Mekjian1991, Comtet2007}.  We can thus
have access to the microcanonical entropy per particle through $s_M =
\ln g(e)/N$.

\begin{figure}[t]
\centering
\includegraphics[width=.85\columnwidth,clip]{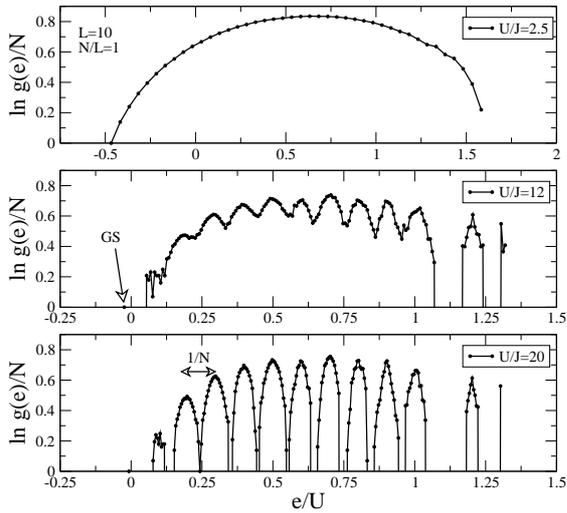}
\caption{Logarithm of the density of states $g(e)$ as a function of
  the energy per particle $e$ in the 1D BHM with density $n=1$ for
  three different interactions. Mott gaps develop at large $U$,
  splitting the density of states into many lobes separated by
  $1/N$.}
\label{fig:density_of_states}
\end{figure}

In Fig.~\ref{fig:density_of_states}, we show the logarithm of the
density of states of the 1D Bose-Hubbard model \emph{on a finite size
  system} ($L=N=10$) for increasing values of the interaction $U$ as a
function of the energy per particle in units of $U$. For small
interactions, the behavior is smooth and one may safely take the
derivative to get the microcanonical temperature.  The system has a
density of states typical of a bound spectrum Hamiltonian, displaying
first positive and then negative temperature regimes. For $U=12J$, in
the Mott phase, one observes a gap to the ground-state in the
low-energy part of the spectrum and also some oscillations over a
typical scale $1/N$. These oscillations are easily understood in the
atomic limit ($J=0$) where they correspond to Mott peaks that have a
high degeneracy, giving this macroscopic density of states at the
center of the lobes. A small $J$ broadens the peaks but the lobes are
expected to survive for large enough $U$ in a finite system, as one
can see for $U=20J$. In this large-$U$ limit, $e_0/U$ gets close to
zero while the maximum energy per site is proportional to the number
of particles (in Fig.~\ref{fig:density_of_states}, the situation at
high energies is a bit different because we cut the maximum number of
bosons onsite). The number of Mott lobes being of order $N^2$, the
density of lobes per unit of $e/U$ grows as $N$ (this remark remains
valid with a cut in the maximum number of bosons per site). This means
that the density of states, as a function of the energy per site, will
be a curve carved into more and more lobes as $N$ increases. For large
enough systems, $\delta e$ will be much larger than the inter-lobe
distance and will pick up the envelope of the lobes as a local density
of states. On finite systems, $\delta e$ and $1/N$ could be of the
same order of magnitude, which makes the definition of the
microcanonical temperature rather difficult since it is very sensitive
to the choice of $\delta e$ and the shape of the peaked distribution.

In the following, the microcanonical ensemble density-matrix $\rho_M$
is defined in the usual way:
\begin{equation}
\rho_M = \sum_{E_{n} \in [\bar{E}-\delta E, \bar{E}+\delta E]} \frac{1}{\Omega} \ket{\psi_{n,f}}\bra{\psi_{n,f}}
\end{equation}
with the ``free'' parameter $\delta E$ as a ``small'' energy window
energy. $\Omega$ is simply the number of eigenstates in the energy
window $[\bar{E}-\delta E, \bar{E}+\delta E]$. The sum over the
eigenstates of $\Ham_f$ must be taken over all symmetry
sectors. Notice that $\delta E$ can be chosen by
hand~\cite{Rigol2008,Rigol2009,Rigol2009a}, or in the same way as the
effective canonical temperature will be: by looking for an approximate
solution of the equation $\bar{E} = \textrm{Tr}[\rho_M \Ham_f]$ (we
recall that $\bar{E} = \elem{\psi_{0,i}}{\Ham_f}{\psi_{0,i}}$ is fixed
by the initial state). In that case, the solution can be multi-valued
so it does not necessarily help. Taking $\delta E$ as the computed
energy fluctuations does not help either because on finite systems,
the distributions for the 1D BHM are quite asymmetric and have large
moments. The choice of $\delta E$ is in general arbitrary and we have
tried to choose the one that gives best results for both the
correlations and the energy. A partial conclusion is that number of
particles required to have a reliable definition of the microcanonical
ensemble can vary a lot depending on the model and the chosen
parameters.  For the 1D BHM, we see that the peculiar shape of the
density of states can be an issue, although it intimately linked to
the physics of the model.

\subsection{Canonical ensemble and effective temperature}

Even though we work on a closed system, we introduce a canonical
density-matrix as it was done in Ref.~\onlinecite{Rossini2009,
 Rigol2009, Rigol2009a} and implicitly in the (grand)-canonical
calculations of Ref.~\onlinecite{Kollath2007}:
\begin{equation}
\label{eq:T_B}
\rho_B = \frac{e^{-\Ham_f / k_B T_B}}{Z}\quad,\text{ with }Z =
\textrm{Tr}[e^{-\Ham_f / k_B T_B}]
\end{equation}
The effective canonical temperature $T_B$ can be defined, as in
Refs.~\onlinecite{Rossini2009, Rigol2009, Rigol2009a}, as the solution of the
equation $\bar{E} = \textrm{Tr}[\rho_B \Ham_f]$. As the mean energy is
a continuous and increasing function of $T_B$, the solution is unique
and the optimization procedure works well. We take $k_B = 1$ in the
following so that temperatures are given in the same units as the
energies. Here again, the trace is taken over all symmetry sectors.
The diagonal ensemble, on the contrary, has non-zero weights only in
the initial state symmetry sector, that is the even parity sector for
the free boson model and the $k=0$ sector in the 1D BHM.  As the
clouds of points of the distributions sometimes look exponential,
another temperature can be defined by fitting the cloud of data with a
normalized Boltzmann law and using a procedure that minimizes the
following cost function between two distributions $\rho_1$ and
$\rho_2$:
\begin{equation*}
\chi(\rho_1,\rho_2) = \sum_{n} (\ln p_{n,1} - \ln p_{n,2})^2 \;.
\end{equation*}
Once convergence is reached, we call $T_D$ the effective temperature
obtained from the distribution.

We lastly recall that provided the density of states scales exponentially
with the energy and the energy fluctuations are negligible in the thermodynamical
limit, the microcanonical and canonical ensembles will lead to the same
thermodynamic functions, and same temperatures.

\subsection{Comparison of observables from different ensembles}

\begin{figure}[t]
\centering
\includegraphics[width=.75\columnwidth,clip]{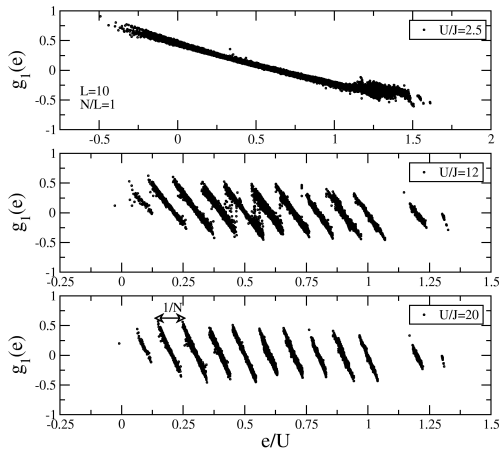}
\caption{Local observable $g_1(e)$ of the 1D BHM as a function of the
  energy per particle for $N=L=10$ and increasing interactions (for
  large quenches, the results were first given in Ref.~\onlinecite{Biroli2009})}
\label{fig:observables_g1}
\end{figure}

\begin{figure}[t]
\centering
\includegraphics[width=.75\columnwidth,clip]{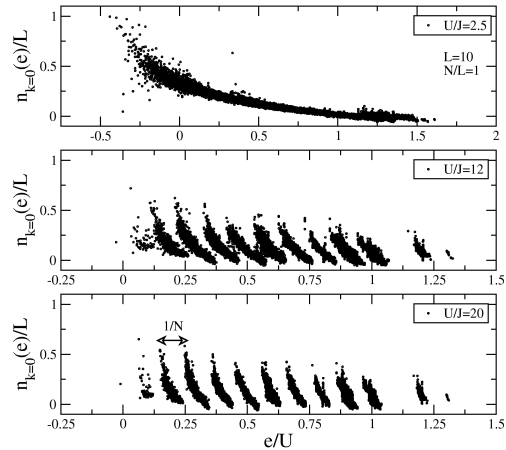}
\caption{Global observable $n_{k=0}(e)$ as a function of the energy
  per particle (same parameters as in Fig.~\ref{fig:observables_g1}).}
\label{fig:observables_nk0}
\end{figure}

\begin{figure}[t]
\centering
\includegraphics[width=.95\columnwidth,clip]{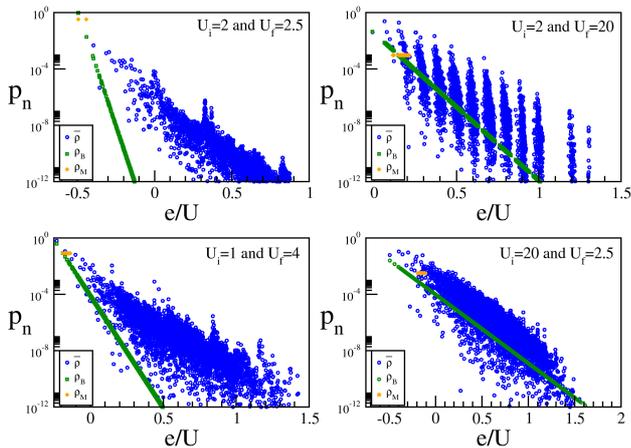}
\caption{(Color online) Comparison of different ensembles for
  different quenches. The effective temperature $T_B$ (given in
  Fig.~\ref{fig:Comparison_gr}) is fixed by the mean energy. The
  results are obtained on a system with $L=N=10$.}
\label{fig:Comparison_ensembles}
\end{figure}

We here focus on the comparison of observables obtained from different
ensembles in the 1D BHM. The evolution of one local and one global
observable as a function of the eigenstates energy per particle is
given in Fig.~\ref{fig:observables_g1} and
\ref{fig:observables_nk0}. Each of these two observables are used
separately in the literature so we here give results for both for
completeness. The observables are the one-particle density-matrix,
defined for a translationally invariant Hamiltonian as:
\begin{equation}
g_r(e) = \frac 1 L \sum_{i=1}^L \elem{\psi_f(e)}{b_{i+r}^{\dag}b_i}{\psi_{f}(e)}\;,
\end{equation}
where $\ket{\psi_f(e)}$ is the eigenstate of energy $e$. $g_r(e)$ is a
local observable since, for a given $r$, it can be attributed to a
subsystem. On the contrary, the momentum distribution $n_k$ integrates
information from all distances and may be considered as a global
quantity:
\begin{equation}
n_k(e) = \sum_{r=-L+1}^{L-1} e^{ikr} g_r(e)\;.
\end{equation}
In Fig.~\ref{fig:observables_g1} and \ref{fig:observables_nk0}, one
observes that both $g_1(e)$ and $n_{k=0}(e)$ evolve smoothly in the
superfluid regime ($U/J=2.5$). One also realizes that the largest
fluctuations are found in the low-energy part of the spectrum,
supporting the energetic argument for the finite size effects. If one
were able to choose $\bar{e}$ in the bulk of the ``superfluid''
spectrum, one would possibly find agreement with ETH. However, for the
finite size systems at hand, one cannot reach the bulk of the spectrum
before the Mott lobes emerge with $\lambda$. As it was shown in
Ref.~\onlinecite{Biroli2009} and here confirmed, the observables
strongly vary within each Mott lobe. We now turn the nature of the
distributions for different quenches and compare the results for $g_r$
obtained by the different
ensembles. Fig.~\ref{fig:Comparison_ensembles} and
\ref{fig:Comparison_gr} gather the data for a small and large quench
from the superfluid region with $U_i=2$.

\subsubsection{Small quench regime in the 1D BHM} 

\begin{figure}[t]
\centering
\includegraphics[width=.95\columnwidth,clip]{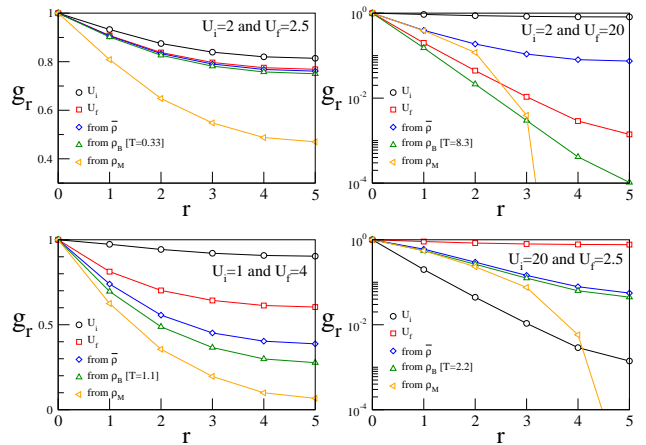}
\caption{(Color online) Comparison of averaged correlations $g_r$
  corresponding the parameters of
  Fig.~\ref{fig:Comparison_ensembles}.}
\label{fig:Comparison_gr}
\end{figure}

\begin{figure}[t]
\centering
\includegraphics[width=.75\columnwidth,clip]{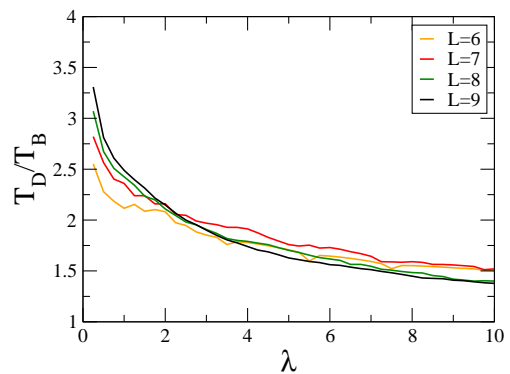}
\caption{(Color online) Ratio of effective canonical temperatures
  obtained from the distribution ($T_D$) and from the mean energy
  ($T_B$). Results are obtained for a quench starting at $U_i=2$.}
\label{fig:ratio_of_T}
\end{figure}

When $U_f=2.5$, the distribution is peaked on the final ground-state
with a large weight $p_0$. The tail displays an exponential-like
behavior that, however, has an effective temperature $T_D$ different
from $T_B$, determined from the energy. This is easily understood from
the fact that only the very few first weights significantly contribute
to the energy, and they are not aligned with the tail. As $\bar{e}$ is
very close to $e_{0,f}$ in this regime and as there are only a very
few energies at the bottom of the spectrum, the microcanonical
ensemble gives a bad mean energy and has only a few number of
eigenstates. In this regime where $p_0$ is close to one, a minimal
microcanonical ensemble would simply be
$\ket{\psi_{0,f}}\bra{\psi_{0,f}}$, although it has no statistical
meaning. Looking at the correlations $g_r$ in
Fig.~\ref{fig:Comparison_gr} shows that they seem to be thermalized in
the sense that $\rho_B$ gives a reasonable account of the
correlations. However, $\ket{\psi_{0,f}}\bra{\psi_{0,f}}$ also gives a
reasonable account for the correlations while $\rho_M$ does not
satisfactorily reproduce them. The system is in a regime dominated by
finite size effects, far below the crossover number of bosons. The
points of Ref.~\onlinecite{Kollath2007} in the ``thermalized'' region
of the phase diagram seem to belong to this regime dominated by finite
size effects. We have also looked at a slightly larger quench
amplitude with $U_i=1$ and $U_f=4$ as in Fig.~3 of
Ref.~\onlinecite{Kollath2007} (however, we work on a slightly smaller
system size and the data displayed in Ref.~\onlinecite{Kollath2007}
were averaged over time, so correlations cannot be quantitatively
compared). Since $p_0$ is smaller, there is a substantial difference
between the correlations in the final ground-state $U_f$ and the one
from the diagonal ensemble. The canonical ensemble still gives the
best agreement with $\bar{\rho}$. In a sense, the shape of the
distributions as given in Ref.~\onlinecite{Roux2009} does explain the
observation of Ref.~\onlinecite{Kollath2007}. Yet, the distribution is
clearly not a true Boltzmann one as the temperature obtained from the
mean energy and other observables are not identical. In order to
investigate this deviation, or difficulty to define an effective
temperature, we have computed the ratio between the two effective
temperatures $T_D$ and $T_B$ in Fig.~\ref{fig:ratio_of_T}. For $L=6$
to 9, it remains between 1 and 3.5 and has a tendency to diverge at
small quenches. Consequently, ETH does not apply here due to the
presence of strong finite size effects, but one cannot claim either
that the system is thermalized even though some correlations look
thermalized in the canonical ensemble. The observed distributions are
specific to this model and to these system lengths and parameters. We
also point out that a similar regime has been observed in
Ref.~\onlinecite{Rigol2009}, corresponding to low effective
temperatures, but for which the diagonal ensemble distributions were
not plotted. 
Still, the behavior of large systems ($N \geq N_c$) in the small
quench regime remains an open but very interesting question as the
low-energy physics will control the behavior. In this respect, we draw
an argument in favor of non-thermalization: for symmetry reasons, the
quench only excites states in the ground-state symmetry sector while
the statistical ensembles average over all symmetry sectors. For
instance, a system with a branch of excitation $E(k)$ can have a $k=0$
gap while the whole spectrum is gapless, hence it could not look
thermalized. Starting from a finite temperature state or including
symmetry breaking terms, like disorder, could partially cure this
symmetry constraint.

\subsubsection{Large quench regime} 

Results for two large quenches at a commensurate density $n=1$, from
the superfluid parameters to deep into the Mott limit and reversely,
are given in Fig.~\ref{fig:Comparison_ensembles} and
Fig.~\ref{fig:Comparison_gr}. For the first one, from $U_i=2$ to
$U_f=20$, the distribution shows very strong fluctuations of the
weights within each Mott lobes~\cite{Roux2009}. In particular, large
weights are present in the low-energy part of the first sub-bands. In
Ref.~\onlinecite{Biroli2009}, it was shown that the larger values of
$g_1$ were correlated to the larger weights (see another example of
such a plot for an incommensurate density in
Fig.~\ref{fig:incommensurate_corr}), explaining that the ETH does not
apply in these finite size systems. This is confirmed by looking at
the time-averaged correlations that are neither reproduced by $\rho_M$
nor by $\rho_B$. DMRG calculations~\cite{Kollath2007, Biroli2009} gave
evidence that a non-thermal correlations $g_r$ survive for system
sizes of order 100.

\begin{figure}[t]
\centering
\includegraphics[width=.85\columnwidth,clip]{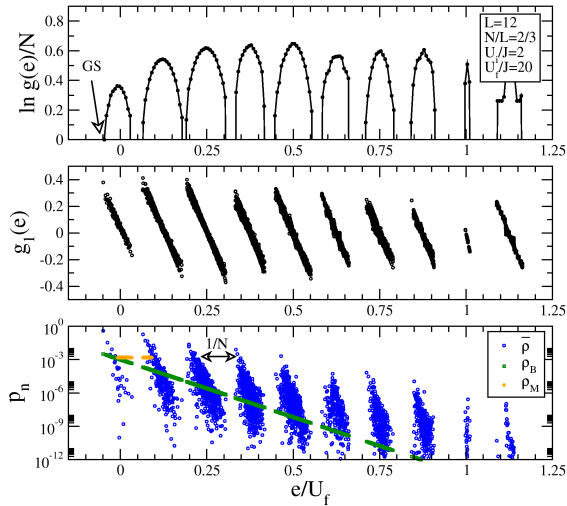}
\caption{(Color online) Quench from $U_i=2$ to $U_f=20$ for an
  \textit{incommensurate} density $n=2/3$ for which there is no
  equilibrium quantum critical point. The structure of the density of
  states, the evolution of the local correlations $g_r$ and the shape
  of the distributions are very similar to the commensurate case.}
\label{fig:incommensurate}
\end{figure}

\begin{figure}[t]
\centering
\includegraphics[width=.85\columnwidth,clip]{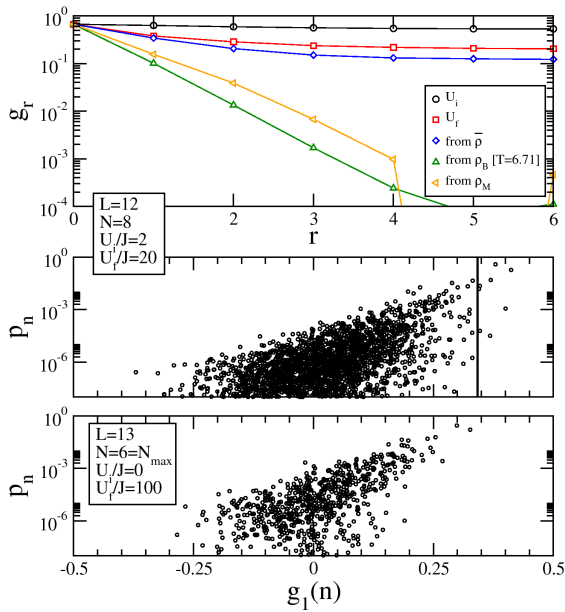}
\caption{(Color online) \emph{Upper panel:} comparison of the
  different observables in different ensembles (same parameters as in
  Fig.~\ref{fig:incommensurate}). \emph{Middle panel}: the $p_n$ vs
  $g_1(n)$ curve gives proof of the non-relaxation towards a thermal
  state for the same parameters as in the upper panel.  \emph{Lower
    panel}: same plot but for the ``integrable'' quench limit $U_i=0$
  and $U_f=100$.}
\label{fig:incommensurate_corr}
\end{figure}

We now elucidate the origin of the observed non-thermalized regime,
first by looking at the effect of the commensurability of the density
in order to determine whether the presence of an equilibrium critical
point plays a role for large quenches. As shown in
Fig.~\ref{fig:incommensurate} and \ref{fig:incommensurate_corr}, the
phenomenology is very similar to the commensurate case with a
non-thermalized regime at large quenches, except that there is no gap
above the ground-state. Quenches that remain in the superfluid region
(data not shown) also have the same behavior as for the commensurate
case. These results suggest that the reason for non-thermalization is
not related to the features of the low-energy spectrum, i.e. to the
presence of a gap above the ground-state, but is related to the
proximity of the $U=\infty$ limit of the model. However, in the small
quench regime where the low-energy part of the spectrum governs the
out-of-equilibrium physics, the opening of a gap can certainly play a
role. Unfortunately, due to the finite size effects discussed in this
paper, this interesting question cannot be addressed with
reliability. For instance, it has been shown recently~\cite{Li2009}
that a quench in the quantum Ising model, which is integrable, is
sensitive to the presence of the critical point. We note that the
lobes could be qualitatively interpreted as stemming from a 
1D gapped single-particle dispersion relation both in the 
commensurate and incommensurate regimes. However, in the latter case, 
there will not be any transition to an insulating state as a function 
of temperature.

One can actually argue that the large-$U$ structure of the
distribution is reminiscent of the atomic limit $U=\infty$ in which we
show that both the weights and the observables fluctuate and are
correlated so that ETH is violated in this limit. What one can show is
that the weights of a quench from $U_i=0$ to $U_f=\infty$ depend on
the configuration in each of the degenerated Mott peaks of the
$U_f=\infty$ limit. This argument does not rely on the $n=1$
commensurability condition. Indeed, the eigenstates of the final
Hamiltonian are simply the set of configurations $\{n_j\}_{j=1,L}$
with $n_j$ the onsite occupations. The energy per particle of the
configuration is
\begin{equation*}
\frac{e(\{n_j\})}{U_f} = \frac 1 {2N} \sum_{i=j}^L n_j(n_j-1)\;.
\end{equation*}
The initial ground-state is the superfluid state that has equal
single-particle probabilities $\mathfrak{p}_j = 1/L$ on each
site. Using formula~(\ref{eq:many-particle-weights}), we get for the
diagonal weights:
\begin{equation}
\label{eq:many-particle-weights-largeU}
p_n = p_{\{n_j\}} = \frac{N!}{n_1!n_2!\cdots n_{L}!} \frac{1}{L^N}\;.
\end{equation}
This makes a connection to the free boson model that we also study,
having the $U_f$ energy spacing between the degenerate levels instead
of $\hbar\omega_f$ and a different energy-configuration relation. The
formula is valid for bare configurations, i.e. when they are not
symmetrized. Using symmetries,
formula~(\ref{eq:many-particle-weights-largeU}) picks up an additional
factor depending on the degeneracy of the generalized Bloch state.
One can see by taking an example of two configurations with the same
energy, or check numerically, that the weights can be different for
configurations with the same energy, in the same way as for the free
boson model. Consequently, in a strongly degenerate Mott peak, the
diagonal weights are not equal and fluctuate. As soon as a
non-integrable perturbation (here the hopping $J$) is turned on and
lifts the degeneracy, the distribution of the weights will still
strongly fluctuate within the Mott lobe.  This explains the findings
of Refs.~\onlinecite{Roux2009,Biroli2009} and of
Fig.~\ref{fig:Comparison_ensembles}.  Another simple observation in
this limit is that two degenerate configurations can have different
expectation values for the observables. An obvious one is the onsite
particle distribution that counts empty, single, double occupations
and so on. The off-diagonal correlation $g_r$ can be non-zero if the
configurations are symmetrized and one can check numerically that they
actually strongly differ for degenerate states. Notice that, in
principle, one has to take into account the off-diagonal part of the
time-averaged density-matrix that is non-zero in this highly
degenerate limit. When one turns on $J$, this off-diagonal part
vanishes and the $g_r$ still fluctuate strongly for eigenstates close
in energy. Lastly, the asymmetrical correlation between the weights
$p_n$ and the observables is also observed in this limit. We show this
numerically on a system with $U_i=0$ and $U_f=100$ in
Fig.~\ref{fig:incommensurate_corr} (we take $U_f/J=100$ and not
$J_f=0$ because one needs a finite, yet very small, $J$ to make
$\bar{\rho}$ diagonal). The numerics for a small $J/U$ in
Fig.~\ref{fig:Comparison_ensembles} and Fig.~\ref{fig:observables_g1}
strongly supports this mechanism as an explanation for the behavior of
both the distributions and the observables. We remark that the
argument works as well for the 2D version of the model that was shown
to have a non-thermalized regime too~\cite{Kollath2007}. The fate of
this explanation in the thermodynamical limit is yet an open
question. A scenario could be that this mechanism works above a
certain critical quench amplitude $\lambda_c(N)$ but how this critical
value behaves as $N\rightarrow \infty$ remains a difficult
question. Consequently, one may understand the finite size effects
stemming from the large-$U$ limit as another $N_c(\lambda)$ line in
Fig.~\ref{fig:BHM_Nc} increasing with $\lambda$ and that is specific
to this model. Yet, non-thermalization in the thermodynamical limit in
the BHM cannot be excluded as well. Experiments in cold
atoms~\cite{Greiner2002} work with a relatively small number of atoms
and can easily reach this large-$U$ limit so that such considerations
are physically relevant.

We also give results for a quench from the Mott to the superfluid
limit. There, one could expect from Fig.~\ref{fig:observables_g1} and
~\ref{fig:observables_nk0} that ETH could work since the observables
behave smoothly with $e$ in the final Hamiltonian. However, for the
accessible sizes, one observes that the Boltzmann law still work
better than the microcanonical ensemble, with large weights at low
energies. We conclude that the breakdown of the ETH could here be
attributed to finite size effects.

\subsubsection{Free boson model} 

\begin{figure}[t]
\centering
\includegraphics[width=.75\columnwidth,clip]{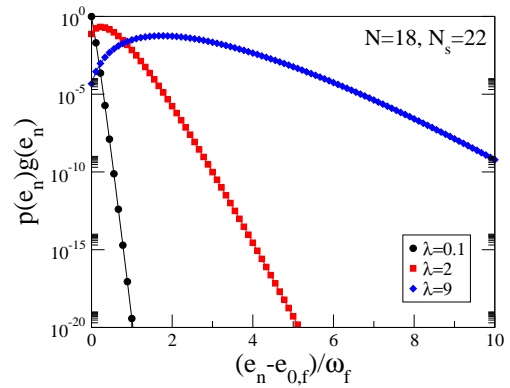}
\caption{(Color online) Evolution of the distribution times the
  density of states of the diagonal ensemble for free bosons in an
  harmonic trap as a function of the quench amplitude. Here, the $p_n$
  are the sum of the diagonal weights in each highly degenerate
  excitation sector.}
\label{fig:distrib_free_bosons}
\end{figure}

\begin{figure}[t]
\centering
\includegraphics[width=.75\columnwidth,clip]{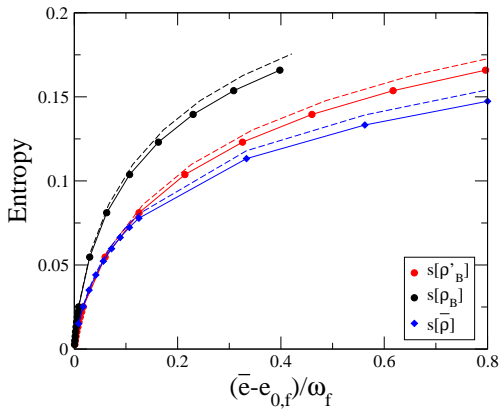}
\caption{(Color online) Von Neumann entropy per particle versus energy
  for the free boson model. Dashed lines are data for $N=17$, full
  lines for $N=18$ ($N_s=22$).}
\label{fig:entropie}
\end{figure}

We now briefly discuss the evolution of the distribution for the free
boson model for a fixed number of bosons and increasing
$\lambda$. Very surprisingly, the distribution of the single-particle
weights versus single-particle energies $\varepsilon_{2n} = 2n \hbar
\omega_f + \hbar \omega_f / 2$ has some remarkable features (we recall
that only the even levels can be occupied from symmetry reasons). In
the limit of large energy $\varepsilon \sim 2n$, we have
\begin{equation*}
  \mathfrak{p}_{2n} \simeq \mathfrak{p}_{0}(\lambda ) \frac{e^{-2n \ln\md{\frac{\lambda+2}{\lambda}}}}{\sqrt{\pi 2n}}
  \propto \frac{e^{- \varepsilon/T(\lambda)}}{\sqrt{\varepsilon}}
\end{equation*}
which has an exponential tail with the effective temperature
$T(\lambda) = \hbar\omega_f / \ln\Md{\frac{\lambda+2}{\lambda}}$. In
the limit of small quenches, the distribution is Boltzmann-like with a
temperature $T(\lambda) \simeq -\hbar\omega_f / \ln|\lambda/2|$ going
to zero. This exponential-like behavior is not generic and a simple
counter-example can be found in the case of an expanding
box~\cite{Aslangul2008}. For the many-particle situation with $N=18$
and $N_s=22$, we give in Fig.~\ref{fig:distrib_free_bosons} the
evolution of the distribution for increasing $\lambda$. For small
quench, the behavior looks like Boltzmann (we do not expect a pure
exponential law due to the presence of the degeneracy function $g(e)$,
see below for a quantitative comparison) and it can be understood from
the fact that the main contribution comes from single-boson
excitations that have the same weights as the single-particle ones.
When $\lambda$ is increased, the energy distribution gets peaked around a
low-energy level and is strongly anisotropic with the maximum at a
different place from the mean energy. This distribution finally develops
a high-energy tail for large $\lambda$. One can compute analytically
the third moment $M_3 = {\rm Tr}(\bar{\rho} (\Ham-\bar{E})^3)$ which
is non-zero and scales as $N$ showing that the distribution remains
anisotropic and that the anisotropy $(M_3)^{1/3}/(\bar{E}-E_{0,f})$
decreases as $N^{-2/3}$. In order to compare the distributions from
different ensembles, we use the von-Neumann entropy of a
density-matrix $\rho$ which is defined as $S_{vN}(\rho) =
-\textrm{Tr}[ \rho \ln \rho]$. Contrary to observables, this quantity
is more sensitive to the tail of the distribution. $S_{vN}/N$ for the
Boltzmann and diagonal ensembles are shown in
Fig.~\ref{fig:entropie}. The density-matrix $\rho_B'$ is a Boltzmann
distribution but restricted to the even parity levels only. We see
that for small quenches, $s(\rho_B')$ and $s(\bar{\rho})$ are very
close. The larger entropy for $s(\rho_B)$ is simply due to the fact
that half of the Hilbert space is not accessible to $\bar{\rho}$ for
symmetry reasons: $s(\rho_B')$ and $s(\rho_B)$ are actually the same
up to a factor 2 in the energy. Comparing the data to the
microcanonical entropy is not relevant here because of finite size
effects (energy discretization and small degeneracy of the first
levels) for the values of the mean energy accessible here.

\section{Conclusions}

The first conclusion we would like to highlight is that, when carrying
out numerical simulations on a finite system, one has to care both
about the quench amplitude and the size of the system to see in which
region of the spectrum are the main weights of the diagonal ensemble
distribution. It has been shown that, although the low-energy part of
the spectrum is the place where the most interesting physics is
expected, one experiences large finite size effects when exploring
it. A crossover number of particles, distinguishing between the small
quench regime and the large quench regime, can be tentatively defined
from energetic considerations or from the static fidelity between the
ground-states of the initial and final Hamiltonians. One advantage is
that they can be computed numerically with few finite size effects
(for the energy based criteria) or with ground-state techniques that
work on larger systems (for both criteria). The numbers have been
computed for the two models under study. As the system follows a
finite size crossover between the two regimes, it can actually happen
to be difficult for nowadays numerics to be close enough to the
thermodynamical limit, where ETH is expected to work generically, even
though some examples can be found in the
literature~\cite{Rigol2008}. This actually is what happens for the
one-dimensional Bose-Hubbard model as we have seen.  Hence, the
thermalization-like regime in the small quench limit deduced from
observables comparison and the qualitative Boltzmann-like structure of
the distribution cannot be considered as truly thermalized because of
dominant finite size effects. Furthermore, sizes accessible with full
diagonalization cannot reach the bulk of the energy spectrum before
the structure of the spectrum resembles the infinite-$U$ atomic limit.
The free boson model nicely illustrates the crossover from a
Boltzmann-like distribution, up to phase space constraints, at small
quench to a different distribution. We note that due to the large
density of states and to negligible energy fluctations, we may expect
the quench, canonical and microcanonical distributions to eventually
be equivalent in the thermodynamical limit. However, we have discussed
the fact that the smaller the mean-energy (or equivalently the
temperature), the larger are finite-size effects. We do not believe
that the observed finite-size and canonical-like distributions at
small quenches are generic (notice that no claim in that direction was
made in Ref.~\onlinecite{Roux2009}) and they may better be simply
understood as (counter-)examples.

The second important conclusion is that we have shown that the
non-thermalized regime observed \emph{on finite systems} for large
quenches in the 1D Bose-Hubbard model is actually related to the
proximity of the $U=\infty$ atomic limit, something that may
qualitatively be equivalent to the proximity of an integrable
point. Indeed, this regime does not depend on the low-energy features
at a commensurate density, i.e. to the presence of the superfluid-Mott
transition, and besides, the structure of the diagonal ensemble stems
from the $U=0\rightarrow U_f=\infty$ quench limit of the Bose-Hubbard
model. In non-integrable models, the challenging issues on what are
the features of the small quench regime for very large sizes and how
the non-thermalized regime neighboring an integrable point survive in
the thermodynamical regime seem to be hardly accessible to current
numerical algorithms.

\acknowledgments

I thank T. Barthel, P. Calabrese, D. Delande, F. Heidrich-Meisner,
C. Kollath, A.M. L\"{a}uchli, P. Leb{\oe}uf, A. Polkovnikov and
D. Ullmo for fruitful discussions. I particularly thank T. Barthel and
F. Heidrich-Meisner for pertinent comments on the manuscript. I am
indebted to M. Rigol~\cite{Rigol2009b} for pointing out that the
temperature extracted from the distribution $T_D$ can be different
from the one defined from the mean energy $T_B$.

\end{document}